\documentclass[english,aps,twocolumn,prl,showpacs,superscriptaddress]{revtex4-1}
\usepackage[T1]{fontenc}
\usepackage[latin9]{inputenc}
\usepackage{babel}
\usepackage{graphicx}
\usepackage{bm}
\usepackage{amsmath}
\usepackage{amssymb}
\usepackage{amsfonts}
\usepackage{dcolumn}
\usepackage{bbold}
\usepackage{tikz}
\usepackage[margin=15mm]{geometry}
\colorlet{linkequation}{blue}
\usepackage[colorlinks=true,linkcolor=blue,citecolor=blue,urlcolor=blue]{hyperref}
\usetikzlibrary{shadings}

\usepackage{color}

\begin{document}

\title{Terahertz spin dynamics driven by a field-derivative torque}

\author{Ritwik Mondal}
\email[]{ritwik.mondal@uni-konstanz.de}
\affiliation{Fachbereich Physik, Universit\"at Konstanz, DE-784 57 Konstanz, Germany}
\author{Andreas Donges}
\affiliation{Fachbereich Physik, Universit\"at Konstanz, DE-784 57 Konstanz, Germany}
\author{Ulrike Ritzmann}
\affiliation{Department of Physics and Astronomy, Uppsala University, P.\,O.\ Box 516, SE-751 20 Uppsala, Sweden}
\author{Peter M. Oppeneer}
\affiliation{Department of Physics and Astronomy, Uppsala University, P.\,O.\ Box 516, SE-751 20 Uppsala, Sweden}
\author{Ulrich Nowak}
\affiliation{Fachbereich Physik, Universit\"at Konstanz, DE-784 57 Konstanz, Germany}

\begin{abstract}
Efficient manipulation of magnetization at ultrashort time scales is of particular interest for future technology. 
Here, we numerically investigate the influence of the so-called field-derivative torque, which was derived earlier based on relativistic Dirac theory  [Mondal {\it et al.}, Phys.\ Rev.\ B {\bf 94}, 144419 (2016)], on the spin dynamics triggered by ultrashort laser pulses. We find that only considering the THz Zeeman field can underestimate the spin excitation in antiferromagnetic oxide systems as, e.g., NiO and CoO. However, accounting for both, the THz Zeeman torque and the field-derivative torque, the amplitude of the spin excitation increases significantly. Studying the damping dependence of field-derivative torque we observe larger effects for materials having larger damping constants.

\end{abstract}

\maketitle

{\it Introduction:}
 The excitation of magnetic materials with THz pulses has recently come into the focus of current research as a possible route towards terahertz spintronics \cite{Kampfrath2010,Kampfrath2013,Baierl2016PRL,Pashkin2013,Wienholdt2012PRL,Baierl2016NatPhoto,Donges2018}. In these experiments a THz laser pump is employed to drive a high-frequency spin wave excitation after which its decay is monitored using magneto-optical spectroscopy with a near-infrared laser. 
 These experiments show that the  THz pulses can {\it coherently} manipulate  the spin degrees of freedom on the picosecond timescale.
  The explanation of these observations requires dedicated spin-dynamics simulations that are capable to treat accurately the action of the short excitation pulse ($\le 1$ ps) as well as both the fast and slow responses of the spin system.

Magnetization dynamics can very efficiently be described by the Landau-Lifshitz-Gilbert (LLG) equation of motion. The  LLG equation has traditionally two parts: precession of the magnetization vector around an effective field and a transverse damping of it \cite{landau35,Gilbert1955,gilbert04}. The LLG equation conserves the magnitude of the magnetization vector and it  has been employed to describe magnetization dynamics from macroscopic time scales down to nanoseconds and less  \cite{lakshmanan11,kronmuller2003micromagnetism}. 
For dynamic processes as they are triggered by ultrashort laser pulses, the stochastic extension of the LLG equation with atomistic resolution has been used to describe spin dynamics on time scales down to picoseconds \cite{Kazantseva2007,nowak2007handbook,Evans2014,Hinzke2011PRL,Wieser2011,Skubic2008}. Here, thermal excitations which affect the 
total magnetization are taken into account via thermal fluctuations in the spin system. However, at those ultrashort timescales relativistic and ultra-relativistic quantum effects can become relevant \cite{Henk2012,Mondal2017Nutation,MondalJPCM2018}.

In an earlier work, starting from  the relativistic Dirac Hamiltonian that included the magnetic exchange interactions, we rigorously derived the LLG equation of motion \cite{Mondal2016}. In particular, we considered the Dirac-Kohn-Sham equation with external magnetic vector potential and derived an extended Pauli Hamiltonian at the semirelativistic limit using a canonical transformation, the so called Foldy-Wouthuysen transformation \cite{Mondal2015a,Mondal2015b}. Thereby, considering only the field-spin coupling Hamiltonian terms, we derived the precessional and damping motion of a spin moment within LLG spin dynamics. 
Moreover, we have shown that, for a magnetic system driven by a general time-dependent field, the LLG equation has to be supplemented with an additional torque term,  the field-derivative torque (FDT)  which is a relativistic effect as it is connected to spin-orbit coupling.  For the case of a time dependent external magnetic field, this means that not only the Zeeman field couples to the spins but also the time-derivative of the Zeeman field. The quantum derivation of the LLG equation suggests that the corresponding LLG spin dynamics for the unit
 magnetization vector $\bm{m}_i(t)$ has the form \cite{Mondal2016,MondalJPCM2018,Mondal2018PRB}       
\begin{align}
    \frac{\partial\bm{m}_i}{\partial t} & = 
     - \gamma \bm{m}_i\times  \bm{B}^{\rm eff}_i \nonumber\\
    & + \bm{m}_i \times\left[\mathcal{G}\cdot \left( \frac{\partial\bm{m}_i}{\partial t} + \frac{ a^3}{\mu_s} \frac{\partial\bm{H}(t)}{\partial t}\right)\right] \,.
\end{align}
Here, $\gamma$ defines the gyromagnetic ratio, $\mu_s$ is the saturation magnetic moment and $a^3$ is the volume of the unit cell per spin. $\mathcal{G}$ is a tensorial damping parameter that is in general time-dependent and can be computed {\it ab initio}
\cite{Mondal2016}. However, in the following we consider only an isotropic and scalar Gilbert damping parameter, which we denote by $\alpha = \frac{1}{3}{\rm Tr}(\mathcal{G})$.  $\bm{H}(t)$ is the applied Zeeman field that is time-dependent;  $\bm{B}^{\rm eff}_i$ is the effective field, which is the derivative of the total energy with respect to $ \mu_s\bm{m}_i$ as will be specified later on. Note that the effective field, ${\bm B}^{\rm eff}_i$, in Eq. (1) is expressed in Tesla, while the applied Zeeman field, $\bm{H}(t)$, is expressed in Ampere per meter.

In this article, we investigate the effect of the FDT term within atomistic spin dynamics simulations. 
Our preceding analytical calculations showed that the FDT term is explicitly dependent on the damping of the material and the frequency of the applied Zeeman field \cite{Mondal2016,Mondal2018PRB}. If both, frequency and damping are high, we expect the FDT terms to show prominent effects. Consequently, we investigate THz spin dynamics effects with and without FDT in antiferromagnetic oxide systems, which have been studied in several experimental and theoretical works on antiferromagnetic THz spintronics in the past \cite{Kampfrath2010,Baierl2016PRL,Baierl2016NatPhoto,Pashkin2013,Wienholdt2012PRL}, yet without additional FDT terms.
We analyse the damping dependence of the FDT effects to show in which materials one can expect high FDT contributions and we compare simulations at zero and finite temperature to demonstrate that FDT effects are not bound to low temperatures. 

\begin{figure}[t]
\centering
\includegraphics[width=9 cm]{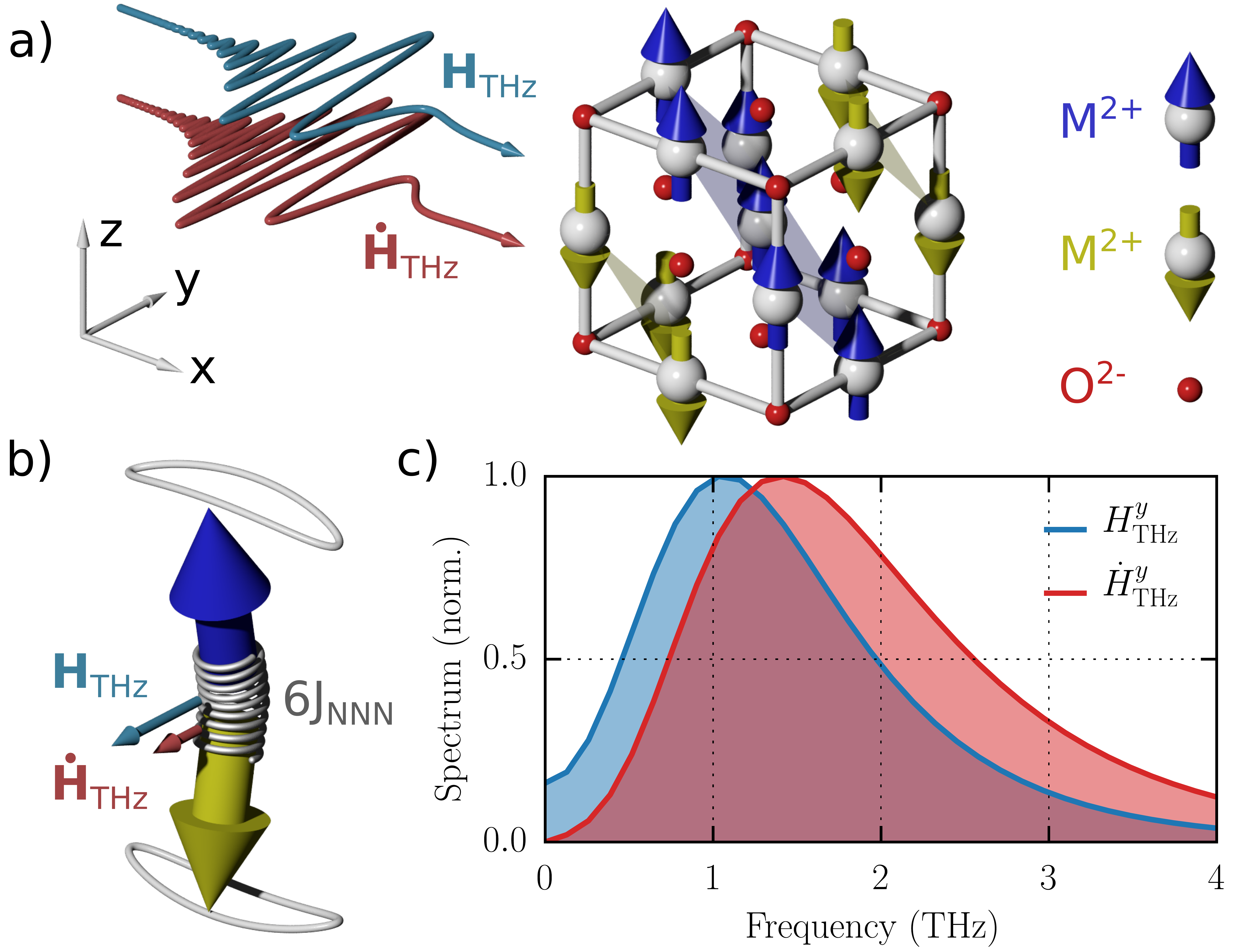}
\caption{(Color Online)   Schematics for the THz excitation of spin moments in an AFM. (a) Ultrashort THz Zeeman pulse  (light blue) and its time derivative (red) acting on fcc MO, where M denotes either Co or Ni. (b) Action of both, the THz pulse and its derivative on the up (dark blue) and down (yellow).  The spring denotes the antiferromagnetic NNN coupling. (c) Fourier transform of both, the THz pulse and its time-derivative (amplitudes normalized to unity).}
\label{THz-pulse}
\end{figure}

{\it Atomistic spin model:}
For the computation of spin dynamics, one often transforms the implicit Gilbert form of the equation of motion to the explicit Landau-Lifshitz form.  This transformation is rather simple for a scalar damping parameter, however, the transformation is cumbersome when a tensorial damping parameter is accounted for \cite{Mondal2016}. In our case, the explicit LLG equation of motion  expanded to take the FDT terms into account can be expressed as
\begin{align}
&\frac{\partial\bm{m}_i(t)}{\partial t} 
 = -\frac{\gamma}{(1+\alpha^2)}\bm{m}_i\times \left( \bm{B}^{\rm eff}_i  - \frac{\alpha a^3}{\gamma\mu_s} \frac{\partial\bm{H}}{\partial t}\right)\nonumber\\
& -  \frac{\gamma\alpha}{(1+\alpha^2)}\bm{m}_i\times \left[\bm{m}_i\times\left( \bm{B}^{\rm eff}_i  - \frac{\alpha a^3}{\gamma\mu_s} \frac{\partial\bm{H}}{\partial t}\right)\right]\,.
\label{LL-equation}
\end{align}
The effective field, $ \bm{B}^{\rm eff}_i$ is calculated as the derivative of total energy with respect to the magnetic moment, i.e., $ \bm{B}^{\rm eff}_i = - \frac{1}{\mu_s}\frac{\partial\mathcal{H}}{\partial \bm{m}_i}$ with $\mathcal{H}$ the total Hamiltonian of the system. 
For the Hamiltonian we consider contributions from the exchange energy ($\mathcal{H}_{\rm exc}$), a crystalline anisotropy energy ($\mathcal{H}_{\rm ani}$) which we restrict here to biaxial, and the Zeeman energy ($\mathcal{H}_{\rm Zeeman}$). These three energies are expressed as 
\begin{align}
    \mathcal{H}_{\rm exc} & = - \sum_{\langle i,j \rangle} J_{\rm NN}^{ij} \bm{m}_i\cdot \bm{m}_j - J_{\rm NNN} \sum_{\langle\langle i,j\rangle \rangle}\bm{m}_i\cdot \bm{m}_j  \,,\\
    \mathcal{H}_{\rm ani} & = - \sum_i \left( d_x m_{ix}^2+d_y m_{iy}^2\right)\,,\\
    \mathcal{H}_{\rm Zeeman} & = - \mu_s \mu_0 \bm{H}(t)\cdot \sum_i\bm{m}_i\,.
\end{align}
$J^{ij}_{\rm NN}$ is the exchange energy constant for nearest neighbours (NN) and $J_{\rm NNN}$ that for next nearest neighbours (NNN).   
$ (d_x,d_y)$ signifies a biaxial anisotropy energy, $\bm{H}(t)$ is the time-dependent Zeeman field and $\mu_0$ the vacuum permeability.  
It is useful to define a Zeeman torque (ZT), $\bm{m}_i \times \bm{H}(t)$, that is induced by the Zeeman field.
In contrast, the FDT terms have been taken into account through the term $\frac{\partial\bm{H}}{\partial t}$ in the LLG equation of motion, Eq. (\ref{LL-equation}). It can be interpreted as an additional contribution to the effective field which acts on the spin system (see Fig.\ \ref{THz-pulse}). 

{\it Application to transition-metal oxides:}
We choose to study the model systems MO (where M denotes either Ni or Co) mainly because  MO supports long-wavelength collective magnon modes at frequencies in the THz regime at room temperature \cite{Kampfrath2010,Kampfrath2013}. Furthermore, it is a comparably simple system which has been studied extensively in previous experimental \cite{Kampfrath2010,Kampfrath2013} and theoretical works \cite{Wienholdt2012PRL}. 
MO has a rock salt structure, where the magnetic M$^{2+}$ ions sit on an fcc lattice [see Fig.\ \ref{THz-pulse}(a)]. It is an AFM with AF2 ground state spin configuration in which alternating (111)-planes couple antiferromagnetically to each other, whereas each (111)-plane is ferromagnetically ordered. Thus each spin has 12 ferromagnetic NN (90$^{\rm o}$ superexchange via O$^{2-}$ atoms) and 6 antiferromagnetic NNN (180$^{\rm o}$ superexchange via O$^{2-}$ atoms) interactions \cite{Hutchings1972,Pashkin2013}. 
The frustrated NN interactions can be divided into two categories: 6 intraplane exchange couplings with $J_{\rm NN}^{\uparrow\uparrow}$ and 6 interplane exchange couplings with $J_{\rm NN}^{\uparrow\downarrow}$ (< $J_{\rm NN}^{\uparrow\uparrow}$), which partially lifts the frustration. 
We use the energy parameters for NiO and CoO systems as described in Table\ \ref{table1} which are obtained from neutron scattering experiments and {\it ab initio} calculations \cite{Hutchings1972,Archer2008,Schron2012}.

The N\'eel temperature, $T_N$, of the NiO spin model is found to be about 340 K (see supplementary material \cite{sm}). This value is notably below the experimentally observed $T_N$ of 540 K \cite{Hutchings1972}, a discrepancy that is known from previous theoretical investigations \cite{Archer2011} and does not affect our conclusions. On the other hand, for CoO, our calculations result in a N\'eel temperature of about 220 K \cite{sm}.
The same N\'eel temperature has been computed from Monte Carlo simulations with the \textit{ab initio} exchange energy parameters for CoO \cite{Archer2008}.
Using these exchange and anisotropy energies, we calculate the antiferromagnetic resonant frequency of our NiO and CoO models as 1.3  and 4 THz, respectively, according to the well-known Kittel formula, $f_{\rm 0} = \gamma  \sqrt{B_{\rm ani}\left(B_{\rm ani} + 2 B_{\rm exc}\right)}$. Here, $ B_{\rm ani}$ and $ B_{\rm exc}$ are the anisotropy and exchange fields, respectively \cite{Kittel1951}. Note that the biaxial anisotropy is considerably larger for CoO as compared to NiO. Due to this higher anisotropy, the resonance frequency is  much higher in the CoO system.

For the two sets of model parameters, we solve the LLG equation [i.e., Eq.\ (\ref{LL-equation})] 
numerically, under the effect of a few cycle THz pulse.  
To treat the thermal fluctuations, we add a stochastic field in terms of white-noise in the effective field $ \bm{B}^{\rm eff}_i$ \cite{nowak2007handbook}. The coupling of the system to the heat bath is quantified by the damping parameter $\alpha$. All the calculations have been performed with a grid of 144$^3$ spins. 
\begin{table}[t]
\caption{Model parameters for exchange and anisotropy energies in meV. The parameters for NiO are taken from Refs.\ \cite{Hutchings1972,Archer2011} and  those for CoO from Refs.\ \cite{Archer2008,Schron2012,text}.}
    \centering
  \renewcommand{\arraystretch}{1.5}
    \begin{ruledtabular}
    \begin{tabular}{c c c c c c}
System & $J_{\rm NN}^{\uparrow\uparrow}$ &$J_{\rm NN}^{\downarrow\uparrow}$  & $J_{\rm NNN}$ &  $d_x$  & $d_y$ \\
\hline
NiO  &   1.4  & 1.3 & -19.0 & -0.10 & -0.005\\ 
CoO  &  1.5 & 1.4 & -12.0 & -2.0 & -0.4 \\ 
  \end{tabular}
 \end{ruledtabular}
\label{table1}
\end{table}      

Terahertz resonant excitation has been studied extensively in NiO and other metal-oxide samples in the linear  as well as nonlinear regime \cite{Kampfrath2010,Kampfrath2013,Baierl2016PRL,Pashkin2013,Baierl2016NatPhoto,Wienholdt2012PRL}. It has been shown that the THz magnetic field can coherently couple to the spins and manipulate the spin degrees of freedom at picosecond timescales. These experimental findings have been compared with the theoretical computation of THz excitation only considering the Zeeman field. Clearly, it can be seen that only the Zeeman field {\it underestimates} the experimental THz excitation in NiO reported in Fig.\ 3 of Ref.\ \cite{Kampfrath2010}. In this work, we employ the THz fields along $y$-direction to the MO spin system at the resonance frequency  (see Fig.\ \ref{THz-pulse}). We model the THz magnetic field by a chirped few cycle pulse of the form
\begin{align}
    H_y(t) = H_0 \cos \left(2\pi f_0 \tau \left[e^{\frac{t}{\tau}} -1\right]\right) e^{-\frac{t^2}{2\sigma^2}}\,,
    \label{THz-pulse-eq}
\end{align}
which is close to the experimental situation \cite{Kampfrath2010}. Here $f_0$ is the resonance magnon frequency, $\sigma$ defines the pulse duration, $\tau$ is the chirped time of the pulse and $H_0$ is the amplitude of the field.  A typical THz pulse and its Fourier transform are shown in Fig.\ \ref{THz-pulse}. As we can see, the NiO resonance frequency we use, 1.3 THz, defines the peak of the Fourier spectrum of the considered THz pulse. 

\begin{figure}[t]
\centering
\includegraphics[scale = 0.64]{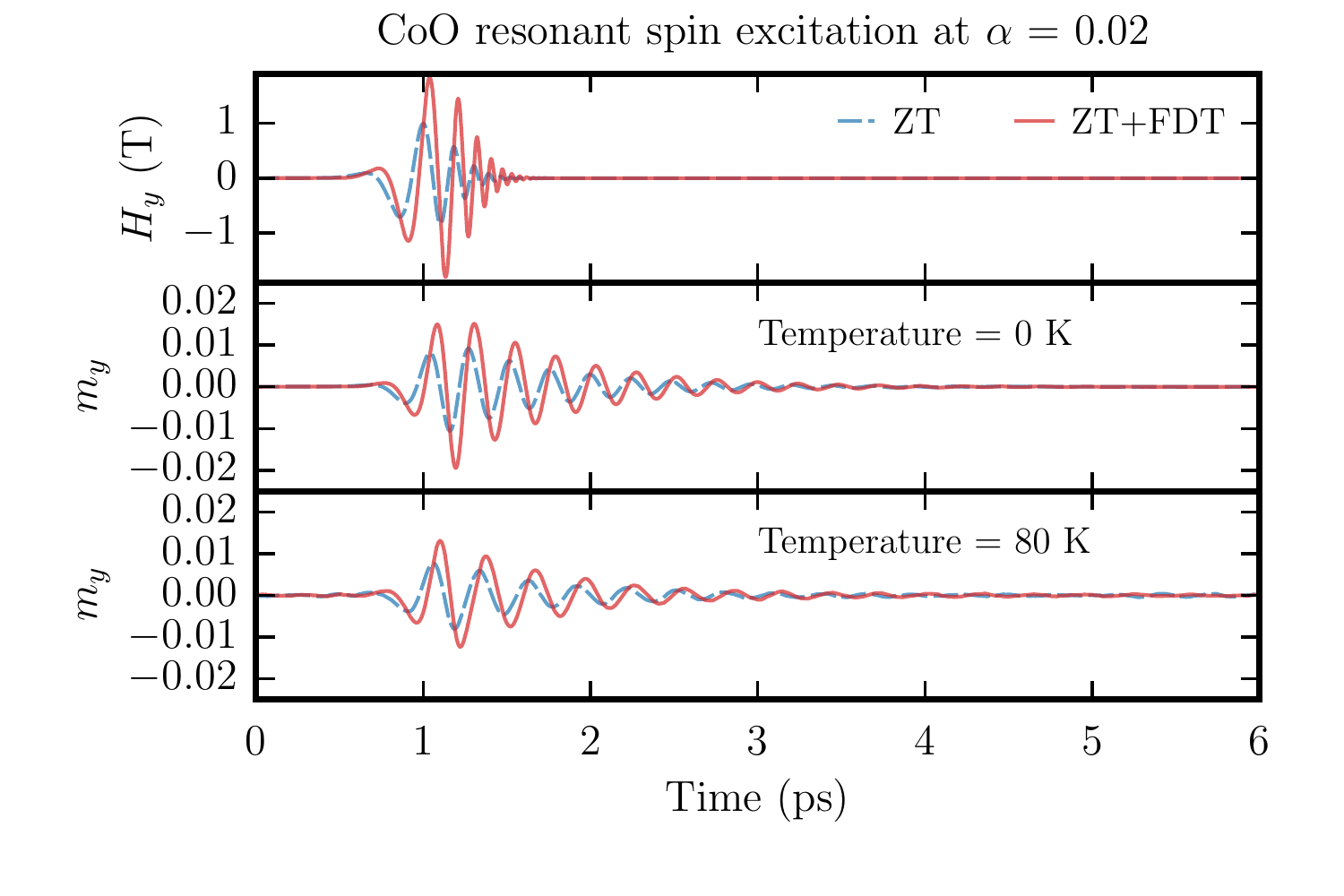}
\caption{(Color Online) THz excitation of the CoO spin system for a damping constant of $\alpha = 0.02$. (Top panel): The THz pulse is shown in blue (dashed line) and the sum of the THz pulse and the FDT effects are shown  by the red line. (Middle panel): Corresponding calculated spin excitation at zero temperature. (Bottom panel): The  calculated spin excitation at a temperature of 80 K.}
\label{CoO-spin-excitation}
\end{figure}

In the following, we will compare the effects of ZT contribution with that of the FDT. Note that in the latter case, one has to consider the ``traditional'' Zeeman effect in addition to the FDT terms as both of them will be present in a time-dependent field. The differentiation introduces a phase shift in the effective field of the FDT  with respect to the traditional Zeeman field, which can be seen in Fig. \ref{CoO-spin-excitation} (top panel). 
In general, the FDT terms will provide a smaller contribution compared to the ZT.  
When comparing the magnitudes of both of the effects, expressed in the effective fields, we find the relationship 
\begin{align}
   \Big \vert \frac{ {\rm FDT} }{  {\rm ZT} } \Big \vert
    & \approx  10^{-10} \, {\rm Sec.} \times \alpha \times f_0\,,
    \label{FDT-compare}
\end{align}
where the values of the model parameters that appear are given in Ref.\ \cite{values}. 
This analysis suggests that the FDT terms due to Zeeman coupling to a field that oscillates with frequency $f_0$ become important for pulses at THz frequencies or higher.  As the FDT is a relativistic spin-orbit effect, it is expected that the effect is stronger at higher damping values which reflects in Eq.\ (\ref{FDT-compare}). Here, in the derivation of FDT, we restrict ourselves to the intrinsic damping parameter of the system, other sources (e.g., spin pumping, radiation etc.) of damping have not been taken into account.

{\it Results and Discussions:} 
We calculate the dynamic spin excitation in NiO and CoO spin models due to THz pumping with and without the FDT terms. The effects of the THz spin excitation at the resonant frequency are shown in Fig.\ \ref{CoO-spin-excitation} for the CoO model. The obtained results show that the THz pulse first excites the resonant magnon mode, which is then eventually damped because of dissipation that is summarized by damping effects with a damping parameter of $\alpha = $ 0.02 in our simulations. This value of the damping is, however, larger than the typical metal-oxide values \cite{Kampfrath2010}.
Note, that when we account for both,  the ZT as well as the FDT terms, the excitation of the magnon mode becomes larger in amplitude. This enhancement in the amplitude is purely because of the FDT terms in the LLG spin dynamics of Eq.\ (\ref{LL-equation}). Therefore, we predict that conventional theory will underestimate
the experimentally observed excitation of resonant THz excitation of the antiferromagnetic magnon mode, 
when only considering the ZT directly, without the additional FDT. 
It also deserves to be noted that the FDT effects appear in the precessional as well as damping terms of the LLG equation of motion in Eq.\ (\ref{LL-equation}), i.e.\
the FDT terms have a field-like and (anti)damping-like contributions. The field-like contribution scales linearly with the damping parameter, however, the (anti)damping-like contribution scales quadratically. For lower damping values, the (anti)damping-like FDT contributions will be much smaller and only the field-like FDT will effectively contribute to the dynamics. In contrast, at higher damping values, both the FDT terms (field-like and damping-like) contribute to the excitation of magnon modes.  

As we pointed out before, the time-derivative introduces a phase shift in the effective field of the FDT term, which, consequently, introduces also a phase shift in the magnon excitation. This effect can also be observed in Fig.\ \ref{CoO-spin-excitation} (middle panel) which displays a phase shift for the excited magnetization oscillation while treating both, the ZT and FDT terms. Furthermore, we investigate the effect of temperature on the spin dynamics in the bottom panel of Fig.\ \ref{CoO-spin-excitation}. Due to the finite temperature (80 K in this case), the magnetic moment is reduced by about 14\% of its initial value for CoO. 
Nevertheless, we observe a very similar behavior in the magnon excitation as in the zero temperature case.
Interestingly, at finite temperature the resonance frequency slightly shifts towards lower values. We find that at 80 K, the resonance frequency decays from 4 THz to 3.6 THz for CoO. This softening of the magnon modes stems from the temperature dependence of the effective magnetic parameters such as the exchange stiffness and the magnetic anisotropy \cite{CALLEN1966,Unai2010}.                                        
Additionally, at finite temperature the maximum spin excitation is decreased with respect to the zero temperature spin excitation \cite{sm}.
Thus, we conclude that even though the FDT terms are smaller than the Zeeman coupling, its effect on the spin system should be observable under realistic conditions. 

\begin{figure}[t]
\centering
\includegraphics[scale = 0.6]{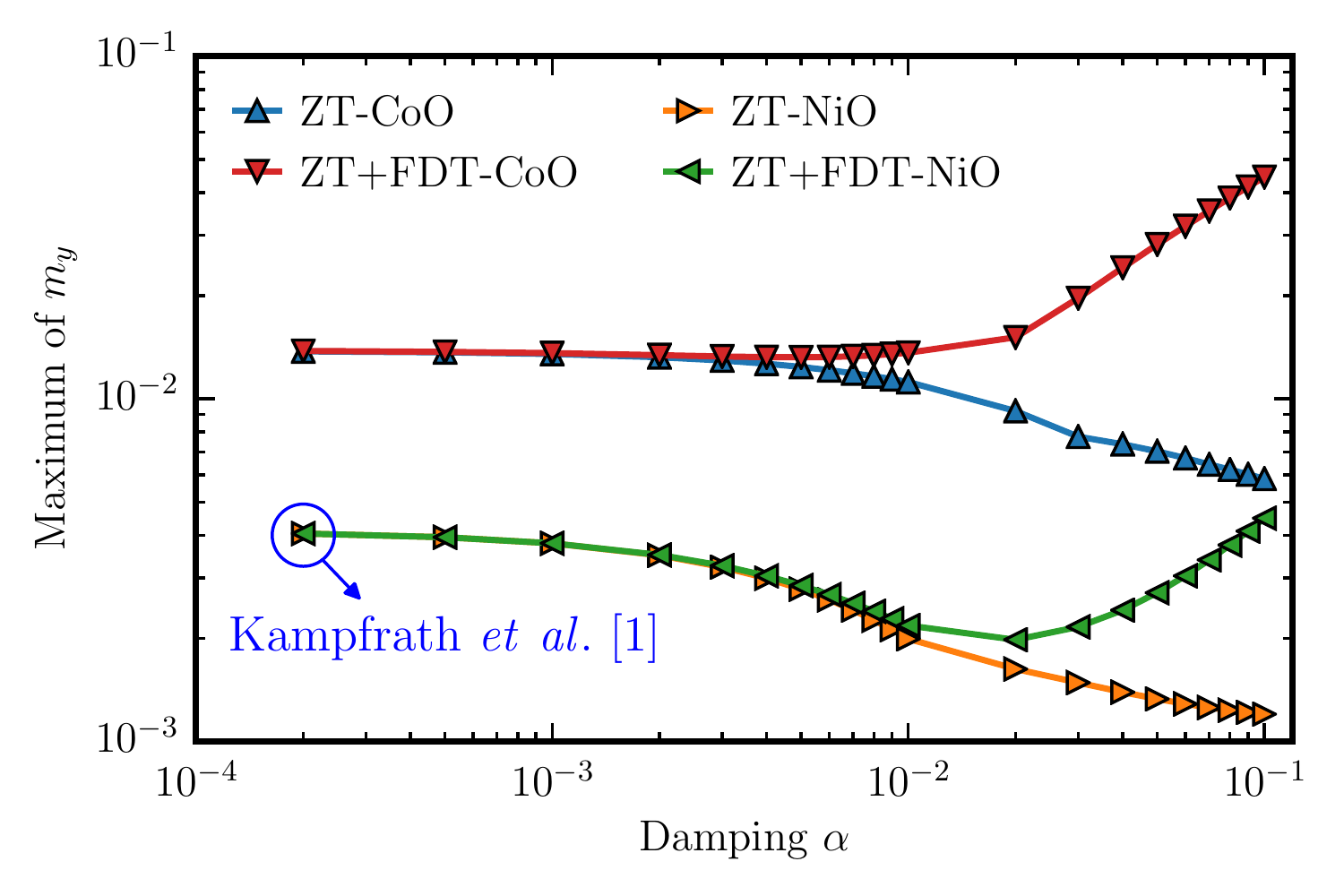}
\caption{(Color Online) Maximum spin excitation as function of the value of the damping parameter at zero temperature. }
\label{damping_vs_spin_excitation}
\end{figure}

In order to forecast the strength of the effect of the FDT for other materials, we study the maximum amplitude of the spin excitation for different 
values of the damping constant, using otherwise the model parameters for NiO and CoO as before. The results are presented in Fig.\ \ref{damping_vs_spin_excitation}. 
Both of the models show a similar behavior though CoO shows an overall higher spin excitation due to its higher resonance frequency. As expected, the FDT terms have a smaller influence for lower values of the damping parameters. For higher damping parameters above 0.01 the influence of the FDT terms become increasingly prominent. 
Here, we observe an ongoing competition between the increasing damping effect and the increasing FDT. While the increasing damping effect tries to lower the maximum spin excitation for the traditional Zeeman coupling, the FDT terms lead to an increasing spin excitation.
In the low damping limit, this competition is won by the damping of the spin excitation which is induced directly by the ZT. As soon as the damping parameter increases above 0.01, the influence of the FDT terms becomes stronger, finally leading to an excitation amplitude which is linear in $\alpha$ due to FDT terms. 

\begin{figure}[t]
\centering
\includegraphics[scale = .6]{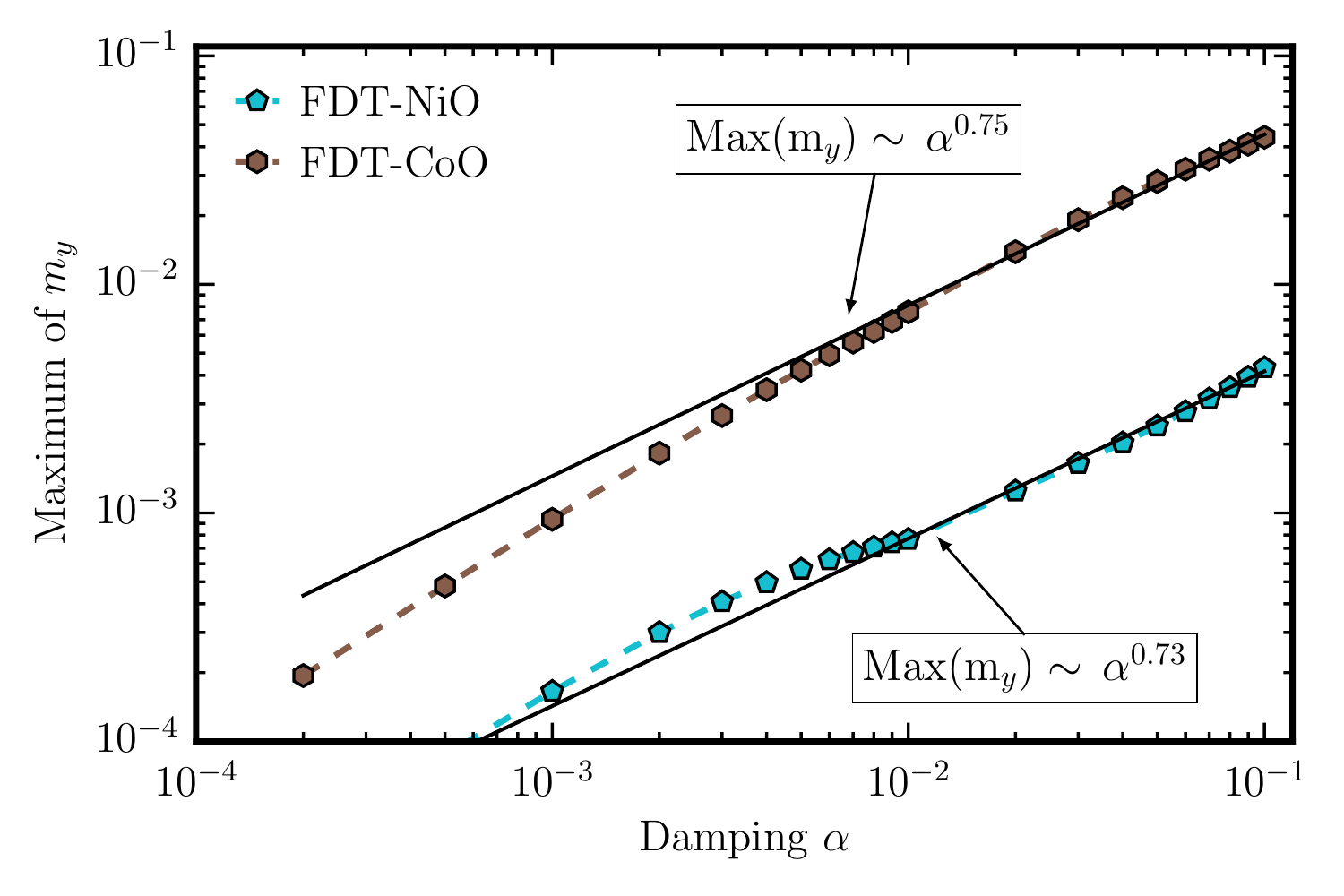}
\caption{(Color Online) Maximum spin excitation as a function of the value of the damping parameters for NiO and CoO, obtained from a simulation with the FDT terms only.}
\label{damping_FDT_NiO-CoO}
\end{figure}

To extract the effect of the FDT terms clearer, we also calculate the maximum spin excitation in simulations where we neglect the Zeeman contribution and take into account the FDT terms only (see Fig.\ \ref{damping_FDT_NiO-CoO}).
Here, a more or less linear behavior is observed even for lower values of the damping constant. 
Moreover, the FDT terms are about 10 times stronger in CoO than NiO due to the higher anisotropy energy in CoO. By fitting the maximum spin excitation  to a functional form max($m_y) \sim \alpha^\delta$, we obtain for CoO and NiO the values $\delta_{\rm CoO} = 0.75$ and $\delta_{\rm NiO} = 0.73$ at higher damping parameters. This means that, for higher damping parameters, both the field-like and damping-like FDT terms  contribute to the dynamics. However, at lower damping parameters we find $\delta_{\rm CoO} = 0.98$ and $\delta_{\rm NiO} = 0.97$ which is consistent with the fact that only field-like FDT terms contribute.  Thus, we conclude that the spin excitations which are induced by FDT terms are similar in antiferromagnetic NiO and CoO.

{\it Conclusions}:
We have examined numerically the consequences of a new extension to the LLG equation of motion which was discovered recently, the FDT \cite{Mondal2016}. From our prior analytical calculations it followed that the FDT is larger for high-frequency excitations and scales linearly with the damping parameter of the system.
We, hence, focused on the effect of the FDT terms on the resonant THz  excitation in NiO and CoO spin systems. 

Due to the FDT terms, the effective time-dependent field becomes asymmetric even if one considers a symmetric THz pulse.
The deformation of the THz pulse due to FDT terms is usually small as compared to the contribution of the Zeeman field. However, the FDT term introduces a phase shift in the effective field pulse. We find that considering only the Zeeman contribution of the THz pulse leads to an underestimation of the coherent spin excitation. In contrast, including the FDT terms in addition to the THz Zeeman contribution in our simulations, we  
observe an increase of the magnon excitation amplitude in NiO as well as CoO. However, the magnitude of FDT-induced magnon excitation for the characteristically low damping constants in the MOs is found to be too small to explain the discrepancy between experimental observations and theory in Ref.\ \cite{Kampfrath2010}.
Nonetheless, we note that a small phase shift is introduced when considering both contributions of the THz pulse, the Zeeman field as well as the FDT terms. These effects are  stronger in CoO as compared to NiO because of its higher resonance frequency. 
We have furthermore studied the damping dependence of the influence of the FDT terms. We find that the FDT terms are more effective for larger values of the damping parameter. Consequently, we expect that the FDT effects would be even stronger for systems with high resonance frequency and larger damping. Such systems include e.g.\  
antiferromagnetic CrPt \cite{Besnus1973,Zhang2012JAP}, where a large spin-orbit coupling results in higher damping and high anisotropy energies, consequently resulting in higher resonance frequencies.  Furthermore, CrPt has higher N\'eel temperature (i.e., stronger exchange coupling which leads to the higher resonant frequency) that is beneficial for room temperature applications. 

We  acknowledge financial support from the Alexander von Humboldt-Stiftung, the Deutsche Forschungsgemeinschaft via RI 2891/1-1 and NO 290/5-1,
and the Swedish Research Council (VR), the Swedish National Infrastructure for Computing (SNIC), and the K.\ and A.\ Wallenberg Foundation (Grant No.\ 2015.0060).

\bibliographystyle{apsrev4-1}

\begin{thebibliography}{37}
\bibitem [{val()}]{values}%
  \BibitemOpen
  \href@noop {} {}\bibinfo {note} {We have used the following set of values for
  calculating the magnitude of FDT terms: the unit cell volume $a^3 = 7.25
  \times 10^{-29}$ m$^3$, the gyromagnetic rattio $\gamma = 1.76\times 10^{11}$
  T$^{-1}$-Sec$^{-1}$, vacuum permeability $\mu_0= 4\pi \times 10^{-7}$
  T-m-A$^{-1}$. These values, together with the saturation magnetization $\mu_s
  = 2.04 \,\mu_{\rm B}$ lead to the FDT prefactor $\frac{a^3}{\gamma
  \mu_0\mu_s} = 1.7\times 10^{-11}$ Sec}\BibitemShut {NoStop}%
\end{thebibliography}
%

\end{document}